\documentclass[aps,prl,twocolumn]{revtex4-1}
\usepackage{amsmath,bm,epsfig,,subfigure}
\usepackage{color}
\usepackage[normalem]{ulem}
\def\Fbox#1{\vskip1ex\hbox to 8.5cm{\hfil\fboxsep0.3cm\fbox{%
  \parbox{8.0cm}{#1}}\hfil}\vskip1ex\noindent}  

\newcommand{\B}[1]{{\bm{#1}}}
\newcommand{\C}[1]{{\mathcal{#1}}}    
\let \= \equiv \let\*\cdot \let\~\widetilde \let\-\overline

\begin{document}
\title{The Glass Transition in Fluids with Magnetic Interactions}
\author{Ricardo Guti\'errez, Bhaskar Sen Gupta, and Itamar Procaccia}
\affiliation{Department of Chemical Physics, The Weizmann Institute of Science, Rehovot 76100, Israel}
\date{\today}
\begin{abstract}
We study the glass transition in fluids where particles are endowed with spins, such that magnetic
and positional degrees of freedom are coupled. Novel results for slowing down in the spin time-correlation
functions are described, and the effects of magnetic fields on the glass transition are studied. Aging
effects in such systems and the corresponding data collapse are presented and discussed.
\end{abstract}
\maketitle

The glass transition occurs when a fluid can be cooled below its melting temperature without crystallizing. When the conditions
are ripe for the creation of such a ``super-cooled" liquid, the temperature can be lowered to the point where the natural relaxation
time of fluctuations in the said liquid begins to exceed our possible observation time, whether experimental or simulational (for a review, see \cite{Cavagna}). At that temperature we speak of the glass transition and denote that temperature as $T_g$. Obviously, to some extent this ``transition"
is in the eyes of the beholder, since having a longer observation time may result in a lower $T_g$ \cite{96EAN}. Experimentally one has
a tremendous variety of glass forming systems, from neat fluids like glycerol \cite{00SD} through mixtures \cite{06Fel}, polymeric liquids \cite{95FD}, metallic glasses \cite{98BBJ},
silicate and water glasses \cite{01VBA} etc. From the point of view of theoretical and simulational research the variety is less impressive, with binary
liquids attracting the majority of effort \cite{91BR,09BSPK}, with a lesser stress on structural examples like water and silicone and even lesser stress on polymeric glasses. An almost neglected but a very interesting type of glass transition is the case in which the positional degrees
of freedom of the participating particles are coupled to magnetic degrees of freedom, for example in metallic glasses in which
one or more of the metallic constituents are also magnetic. Interesting questions like the effect of the existence of a magnetic field
on the glass transition as well as the phenomenon of slowing down of the fluctuations of the magnetic degrees of freedom can be
asked and answered in such fluids. This is indeed the aim of the present Letter.

The example that we chose to simulate was used recently to study the properties the magnetic amorphous solids \cite{Magnetic}.  It is defined by the
Hamiltonian
\begin{equation}
\label{umech}
U(\{\mathbf r_i\},\{\mathbf S_i\}) = U_{\rm mech}(\{\mathbf r_i\}) + U_{\rm mag}(\{\mathbf r_i\},\{\mathbf S_i\})\ .
\end{equation}
 Here $\{\mathbf r_i\}_{i=1}^N$ are the 2-D positions of $N$ particles in an area $L^2$ and $\mathbf S_i$ are spin variables. The mechanical part $U_{\rm mech}$ is chosen to represent a glass forming system with a binary mixture of 65\% particles A and 35\% particles B, with Lennard-Jones potentials having a minimum at positions $\sigma_{AA}=1.17557$, $\sigma_{AB}=1.0$ and $\sigma_{BB}=0.618034$ for the corresponding interacting particles \cite{09BSPK}. These values are chosen to guarantee good glass formation and avoidance of crystallization. The energy parameters chosen are $\epsilon_{AA}=\epsilon_{BB}=0.5$
$\epsilon_{AB}=1.0$, in units for which the Boltzmann constant equals unity. All the potentials are truncated at distance 2.5$\sigma$ with two continuous derivatives. $N_A$ particles A carry spins $\mathbf S_i$; the $N_B$ B particles are not magnetic. Of course $N_A+N_B= N$. We choose the spins $\mathbf S_i$ to be classical $xy$ spins; the orientation of each spin is then given by an angle $\phi_i$ with respect to the direction of the external magnetic field $\B H$ which is along the $x$ axis.

The magnetic contribution to the potential energy takes the form \cite{Magnetic}:
\begin{eqnarray}
&&U_{\rm mag}(\{\mathbf r_i\}, \{\phi_i\}) = - \sum_{<ij>}J(r_{ij}) \cos{(\phi_i-\phi_j)}\nonumber\\&&-  \sum_i K_i\cos^2{(\phi_i-\theta_i(\{\mathbf r_i\}))}-  \mu_A H \sum_i \cos{(\phi_i)} \ .
\label{magU}
\end{eqnarray}
Here $r_{ij}\equiv |\mathbf r_i-\mathbf r_j|$ and the sums are only over the A particles that carry spins. The exchange parameter $J(\mathbf r_{ij})$ is a function of a changing inter-particle position. We choose for concreteness the monotonically decreasing form $J(x) =J_0 f(x)$ where $f(x) \equiv \exp(-x^2/0.4)+D_0+D_2 x^2+D_4 x^4 $ with
$D_0=-6.81\times 10^{-5}\ ,D_2=2.04 \times 10^{-5}\ , D_4=-1.53 \times 10^{-6}$.
This choice cuts off $J(x)$ at $x=2.5$ with two smooth derivatives. Finally, in our case $J_0=3$.
The local axis of anisotropy $\theta_i$ is determined by the local structure. In a super-cooled liquid the structure and the arrangement of particles changes from place to place, and we need to find the local easy axis by taking this arrangement into account. To this aim define  the matrix $\mathbf T_i$:
\begin{equation}
T_i^{\alpha\beta} \equiv \sum_j J( r_{ij})  r_{ij}^\alpha r_{ij}^\beta/\sum_j J( r_{ij}) \ .
\end{equation}
Note that we sum over all the particles that are within the range of $J( r_{ij})$; this catches the arrangement of the local neighborhood of the $i$th particle. The matrix $\mathbf T_i$ has two eigenvalues in 2-dimensions that we denote as $\kappa_{i,1}$ and $\kappa_{i,2}$, $\kappa_{i,1}\ge \kappa_{i,2}$. The eigenvector that belongs to the larger eigenvalue $\kappa_{i,1}$ is denoted by $\hat {\mathbf n}$. The easy axis of anisotropy is given by $\theta_i\equiv \sin^{-1} (|\hat n_y|)$. Finally the coefficient $K_i$ which now changes from particle to particle is defined as
\begin{equation}
\label{KK}
K_i \equiv \tilde C[\sum_j J( r_{ij})]^2 (\kappa_{i,1}-\kappa_{i,2})^2\ ,~~ \tilde C= K_0/J_0\sigma^4_{AB} \ .
\end{equation}
The parameter $K_0$ determines the strength of this random local anisotropy term compared to other terms in the Hamiltonian. For the data shown below we chose $K_0=7.0$. The form given by Eq.~(\ref{KK}) ensures that for an isotropic distribution of particles $K_i=0$. In a super-cooled liquid the direction $\theta_i$ is random.

\begin{figure}
\includegraphics[scale = 0.20]{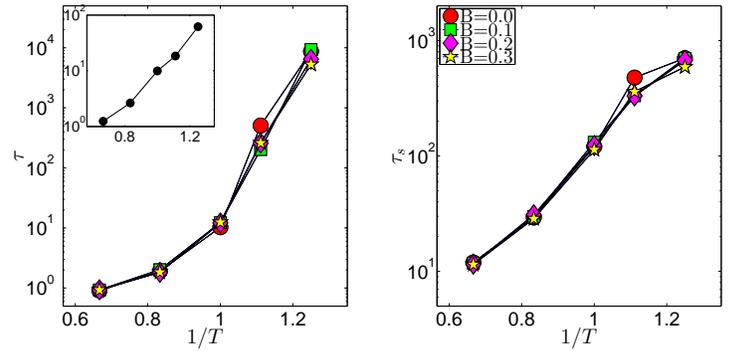}
\caption{ Left panel:The typical relaxation time $\tau$ vs. $1/T$ for a range of external magnetic fields. Inset: a comparison to the relaxation time $\tau$ of the same identical binary mixture without the coupling to the magnetic degrees of freedom. Right panel: typical time $\tau_s$ as a function of $1/T$ for a range of magnetic fields.}
\label{3}
\end{figure}

To simulate this model in molecular dynamics we solved the Newton equations for both positions and spins, taking the mass
of the particles and the moment of inertia of the spin to be unity:
\begin{eqnarray}
\ddot {\B r}_i & =& -\frac{\partial U(\{\mathbf r_i\},\{\phi_i\})}{\partial \B r_i}\ , \label{ddr}\\
\ddot {\phi}_i &=& -\frac{\partial U(\{\mathbf r_i\},\{\phi_i\})}{\partial \phi_i} \ .
\end{eqnarray}
For each chosen temperature the system is initiated from a random configuration of positions and spins, and is equilibrated
in the presence of a chosen value of the magnetic field using the standard Berendsen thermostat \cite{Berendsen}. The range of chosen magnetic
fields from $H=0$ to $H=0.3$ is quite large - at the higher values of $H$ the spins are oriented in the $x$ direction such that the magnetization along the field direction
is above $0.5$ for the lower temperatures considered.  After equilibration the intermediate scattering function and spin correlation function were computed as
\begin{eqnarray}
F_s(\B q,t) &\equiv& \Big\langle \frac{1}{N_A}\sum_{i=1}^{N_A} e^{i \B q\cdot [\B r_i(t) -\B r_i(0)]}\Big\rangle \label{Fs}\\
\C C(t) &\equiv& \Big \langle  \frac{1}{N_A}\sum_{i=1}^{N_A} \frac{\delta \B S_i(t)\cdot \delta \B S_i(0)}
{|\delta \B S_i(0)|^2}  \Big \rangle \ , \label{C}
\end{eqnarray}
where the symbol $\langle \cdots \rangle$ stands for an ensemble average over many (typically 100) initial times $t=0$ (separated
by the relaxation time) and $\delta \B S_i\equiv \B S_i -\langle \B S_i\rangle$. The substraction of the mean spin is obviously
necessary whenever $\B H\ne 0$. The modulus of the wave vector $q=6.08$ represents the first peak of the static structure factor.
\begin{figure}
\includegraphics[scale = 0.20]{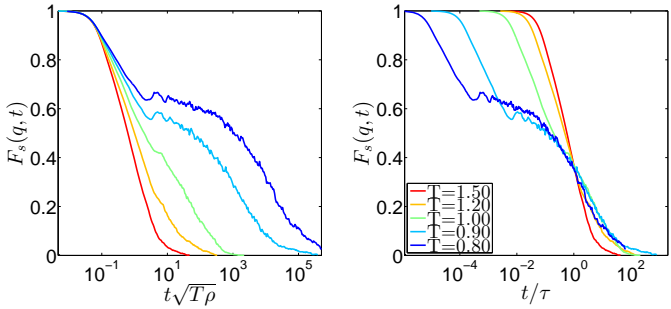}
\caption{ Left panel: The intermediate scattering function vs. time rescaled by the typical cage time for a range of temperatures. Right panel: the same functions vs time rescaled by the $\alpha$ relaxation time $\tau$.}
\label{1}
\end{figure}
\begin{figure}
\includegraphics[scale = 0.20]{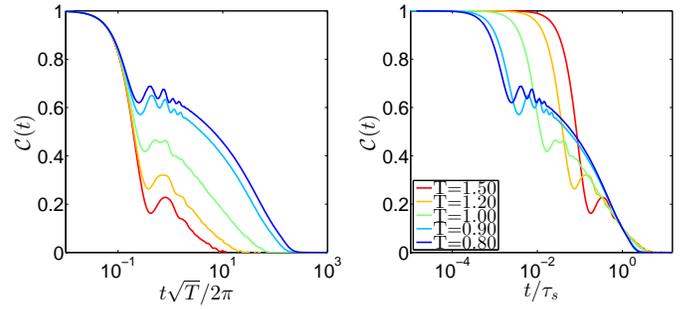}
\caption{Left panel: The spin autocorrelation function vs. time rescaled by the typical spin revolution time for a range of temperatures. Right panel: the same functions vs time rescaled by the relaxation time $\tau_s$.}
\label{2}
\end{figure}
In Figs. \ref{1} and \ref{2} we exhibit the correlation function (\ref{Fs}) and (\ref{C}) as a function of time respectively.
In the left panel the functions are plotted in units of the natural times $\rho^{-1/2}/\sqrt{T}$ and $2\pi/\sqrt{T}$ which
stand respectively for the time to move within the cage for Fig. \ref{1} and the time for a spin to rotate 2$\pi$ radians (since
the velocity is proportional to $\sqrt{T}$). Note the small peak in the correlation functions just after the ballistic regime; this peak has to do with the typical system size and the sound mode bouncing back and forth \cite{94LW}. We have independently checked this identification by changing the system size. The right panel in Fig. \ref{1} represents the same function as on the left panel but with the time rescaled
by the appropriate relaxation time $\tau(T)$ which is defined as the time taken by $F_s(\B q,t)$ to decay from unity to $1/e$.
The right panel in Fig. \ref{2} is shown as a function of time rescaled by $\tau_s(T)$ which is defined as the time taken
for $\C C(t)$ to decay to size 0.1. To our knowledge this is the first simulational result showing the glass transition effect
on the slowing down of the spin autocorrelation function. The typical relaxation time $\tau$ and $\tau_s$ are plotted as a function
of $1/T$ in Fig. \ref{3}.
Examining the data in Fig. \ref{3} we learn that (at least for the parameters used in these simulations) the existence of an external magnetic field has a negligible effect on the dynamical
properties of {\em both} the intermediate scattering function {\em and} the spin autocorrelation function. In fact, while the data shown in Figs. \ref{1} and \ref{2} were presented for zero field, the relaxation functions for any used value of the magnetic field $H$
were quite identical. We find this a surprising
fact that will have to be explained by further theoretical investigations. It is interesting to note at this point that in the range
of simulated temperature the relaxation time $\tau$ spans about 4 orders of magnitude, with a lesser effect on the slowing down
of the spin autocorrelation function where $\tau_s$ ranges only over 3 orders of magnitude. In addition we point out that the
coupling of the positional degrees of freedom to the spins increases the effect of slowing down on the intermediate scattering
function. For comparison we show in the inset in Fig. \ref{3} the relaxation time $\tau$ in the same identical binary mixture
without coupling to the magnetic degrees of freedom. We observe that the coupling added at least one to two orders of magnitude to the
range of slowing down of $\tau$ over the same range of temperatures. In principle one could observe even longer relaxation times
by going to lower temperatures, but for the reasons spelled out above our simulations became too long for $T<0.8$ to be able to provide reliable relaxation functions. In fact for this magnetic system also the numerical integration is slower than for the case of
pure mechanical interactions.

\begin{figure}
\includegraphics[scale = 0.19]{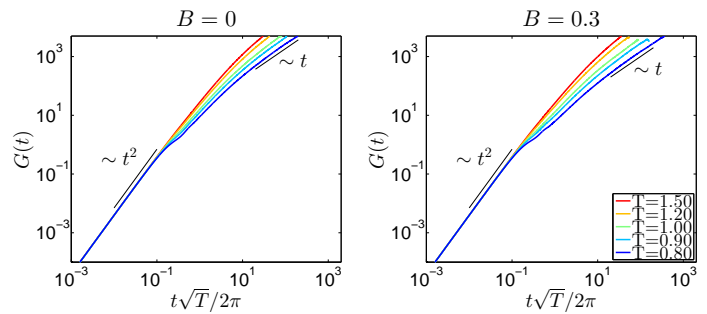}
\caption{The mean-square-rotation of the spins $G(t)$ for two values of the external magnetic field $H=0$ and $H=0.3$.}
\label{4}
\end{figure}

\begin{figure}
\includegraphics[scale = 0.22]{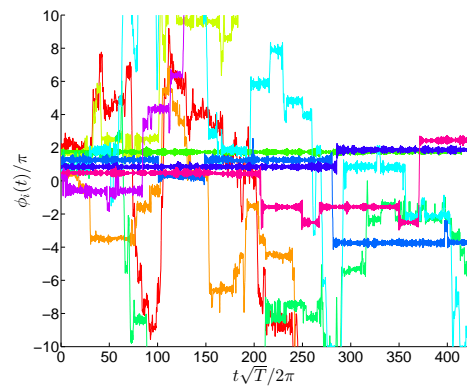}
\caption{Measured angles $\phi_i$ as a function of time for 10 representative spins chosen at random, $H=0$ $T=0.8$.}
\label{5}
\end{figure}

\begin{figure}
\includegraphics[scale = 0.22]{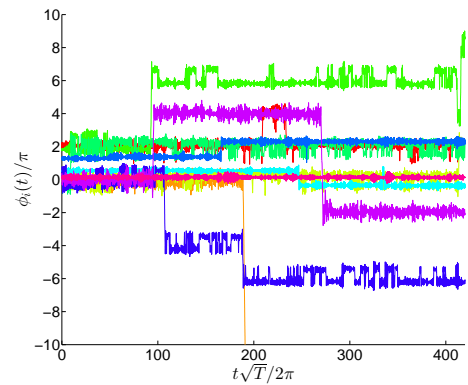}
\caption{Measured angles $\phi_i$ as a function of time for 10 representative spins chosen at random, $H=0.3$  $T=0.8$.}
\label{6}
\end{figure}

Next we present new results on the mean-square-rotation of the spins as a function of temperature and magnetic field.
Define the function $G(t)$ according to
\begin{equation}
G(t)\equiv \frac{1}{N_A}\Big \langle \sum_{i=1}^{N_A} [\phi_i(t) -\phi_i(0)]^2 \Big \rangle .
\end{equation}
At time $t=0$ all the randomly distributed angles $\phi_i$ take on values in the range $[0,2\pi]$. following their dynamics
one observes small fluctuations in the angle punctuated with large excursions, of $\pi$, $2\pi$ or sometime more.
Naively one would expect that with $H=0$ this quantity would grow without limit, possibly proportional to $t$ as is expected in
diffusive motion, but that for a high value of $H$ it would saturate since only small fluctuations around $\phi_i=0$ are expected.
In fact this is not the case. In Fig. \ref{4} one can see that this function appears to grow without limit for both
$H=0$ and $H=0.3$.
The function grows ``ballistically fast", proportional to $t^2$ for short times. When the temperature is high this behavior
continues for a longer time before crossing over to a slower increase, which is eventually diffusive, proportional to $t$.
The cross over to diffusive behavior is enhanced for high field as compared to $H=0$. To understand why this function
continues to grow in time even for $H=0.3$ we show in Figs. \ref{5} and \ref{6} the measured angle of 10 representative randomly chose spins as a function of time.
One sees that when $H=0$ the spins can make large rotations quite freely, explaining the extended region of ballistic behavior.
For the large field $H=0.3$ one sees the expected small fluctuations around the preferred direction of the field, but there are
also unexpected $2\pi$ rotations every now and then which explain the unbounded increase of $G(t)$ even when the field is high.

\begin{figure}
\includegraphics[scale = 0.20]{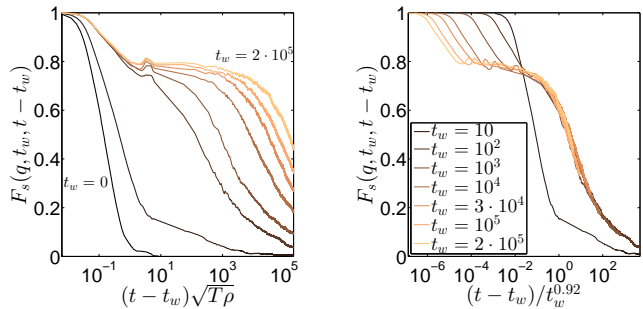}
\caption{ Left panel: the correlation function $F_s(t-t_w)$ as a function $t-t_w$ normalized by the typical cage time, for a range of
$t_w$ between 0 and $10^5$, see the right panel for all the used values of $t_w$. Right panel: the same function plotted against
a rescaled time $(t-t_w)/t_w^{0.92}$.}
\label{7}
\end{figure}

\begin{figure}
\includegraphics[scale = 0.20]{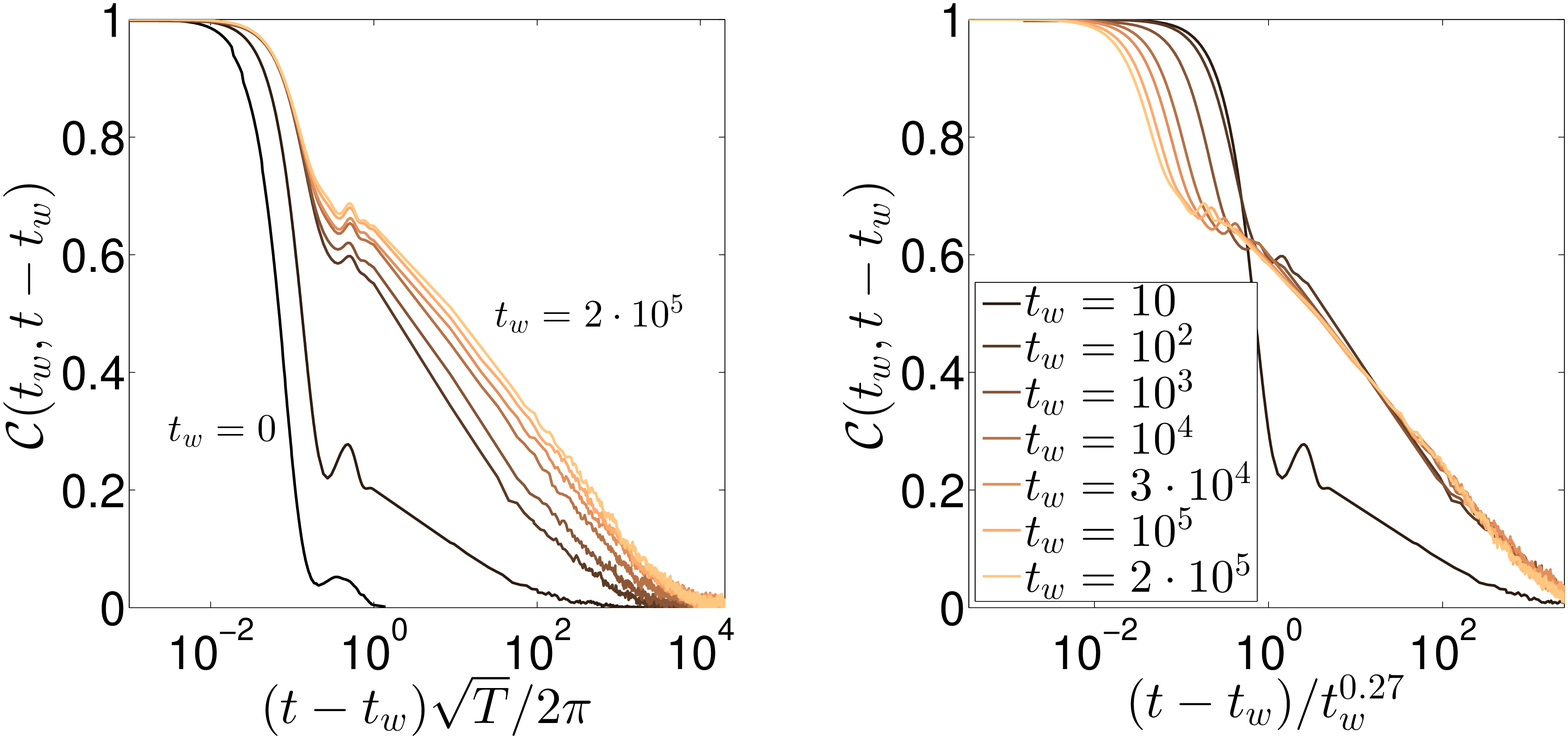}
\caption{ Left panel: the correlation function $\C C(t-t_w)$ as a function $t-t_w$ normalized by the typical spin revolution time, for a range of $t_w$ between 0 and $10^5$, see the right panel for all the used values of $t_w$. Right panel: the same function plotted against a rescaled time $(t-t_w)/t_w^{0.27}$.}
\label{8}
\end{figure}

Finally we study the properties of this system under aging  \cite{KobBarrat}. To this aim we prepared an equilibrated system at $T=3$ and quenched
it suddenly (at $t=0$) to $T=0.4$. At this point we waited a waiting time $t_w$ before starting to measure both correlation functions.
The results of this study are summarized in Fig. \ref{7} and \ref{8}.\\
As expected, the relaxation functions become slower and slower as the waiting time increases, but the positional and the spin
relaxation time do it with very different rates. This is stressed by the collapsed data in the right panels of the two
figures, where we show that the relaxation time of these function scale themselves as $t_w^\eta$, with very different $\eta=0.92$
for $F_s$ and $\eta=0.27$ for $\C C$. The different exponents reflect what was found before, i.e. that the glass transition
has a stronger effect on the positional relaxation function and a weaker effect on the spin autocorrelation function. Yet the actual
number that these exponent take on call for further theoretical studies that are beyond the scope of this Letter.

In summary, we have presented results on the dynamics of the glass transition in a model glass in which the positional
degrees of freedom are coupled to magnetic ones. We have noted a significant effect on the slowing down of the positional degrees of freedom and a novel slowing down in the fluctuations of the spin degrees of freedom. Questions like the existence of a typical length and the degree of fragility of the present model will be discussed elsewhere. Finally we note that the model chosen in not unique, and further study along these lines is necessary to explore the interesting consequences of such couplings. In the coming future we will present additional models which will afford a variety of new effects that will require additional theoretical considerations.

\acknowledgments

This work had been supported in part by the ERC under the ``ideas" grant STANPAS. We thank George Hentschel for many illuminating
discussions regarding how to model magnetic glasses.

\end{document}